\documentclass[5p,twocolumn,times,number]{elsarticle}

\usepackage{graphicx}
\usepackage{amsmath}   
\usepackage{hyperref}

\begin{document}

\begin{frontmatter}

\title{Development of cryogenic calorimeters to measure the spectral shape of $^{115}$In $\beta$-decay}

\author[GSSI,LNGS]{E.~Celi}
\author[Leibniz]{Z.~Galazka}
\author[LNGS]{M.~Laubenstein}
\author[QUEEN,MCINST]{S.~Nagorny}
\author[LNGS]{L.~Pagnanini}
\author[LNGS]{S.~Pirro}
\author[GSSI,LNGS]{A.~Puiu}

\address[GSSI]{Gran Sasso Science Institute, 67100, L'Aquila, Italy}
\address[LNGS]{INFN  Laboratori Nazionali del Gran Sasso, I-67100 Assergi (AQ), Italy}
\address[Leibniz]{Leibniz-Institut für Kristallzüchtung, Max-Born-Str. 2, 12489 Berlin, Germany}
\address[QUEEN]{Department of Physics, Engineering Physics and Astronomy,Queen’s University, Kingston, ON, K7L 3N6, Canada}
\address[MCINST]{Arthur B. McDonald Canadian Astroparticle Physics Research Institute, Department of Physics, Engineering Physics and Astronomy,Queen’s University, Kingston ON K7L 3N6, Canada}

\begin{abstract}
The spectral shape of forbidden beta-decays is a crucial benchmark for nuclear physics calculations and has important implications also for astroparticle physics experiments. Among the interesting isotopes in this field, $^{115}$In is an excellent candidate, being a primordial nuclide featuring a good compromise among Q-value (497.954~keV) and half-life ($4.41 \times 10^{14}$ yr). In this paper, we propose to exploit a cryogenic calorimeter based on In$_2$O$_3$ crystal to perform a high-precision measurement of the $\beta$-decay energy spectrum, discussing also the results obtained within a preliminary test of this crystal and the next steps to improve the detector performance.
\end{abstract}

\end{frontmatter}
\section{Introduction}\label{Sec:Introduction}
The naturally occurring $^{115}$In is one of three isotopes, together with $^{113}$Cd and $^{50}$V, for which a fourth-forbidden non-unique $\beta$-decay may occur, i.e. the nucleus parity does not change while the angular momentum changes by four units, in symbols $\Delta L^{\Delta P} = 4^{+}$. 

$^{113}$Cd is the most studied isotope, thanks to well established detection technologies. In particular, it was measured using CdWO$_4$ crystals by the KINR-DAMA collaboration~\cite{Belli:2007zza} and CdZnTe by the COBRA experiment~\cite{Bodenstein-Dresler:2018dwb}, both with high performance but limited by the high energy threshold (CdWO$_4$), or by presence of other active isotopes in the detector (CdZnTe), which spoil the result on $^{113}$Cd.

$^{50}$V presents an opposite situation, being its $\beta$-decay not yet observed. Different experimental approaches are under development~\cite{Laubenstein:2018euc,Pattavina:2018nhk}
to investigate such decay beyond the current limit, i.e. $T_{1/2} = 1.9 \times 10^{19}$ yr~\cite{Laubenstein:2018euc}.

\begin{figure}[!t]
    \centering
    \includegraphics[width=0.5\textwidth]{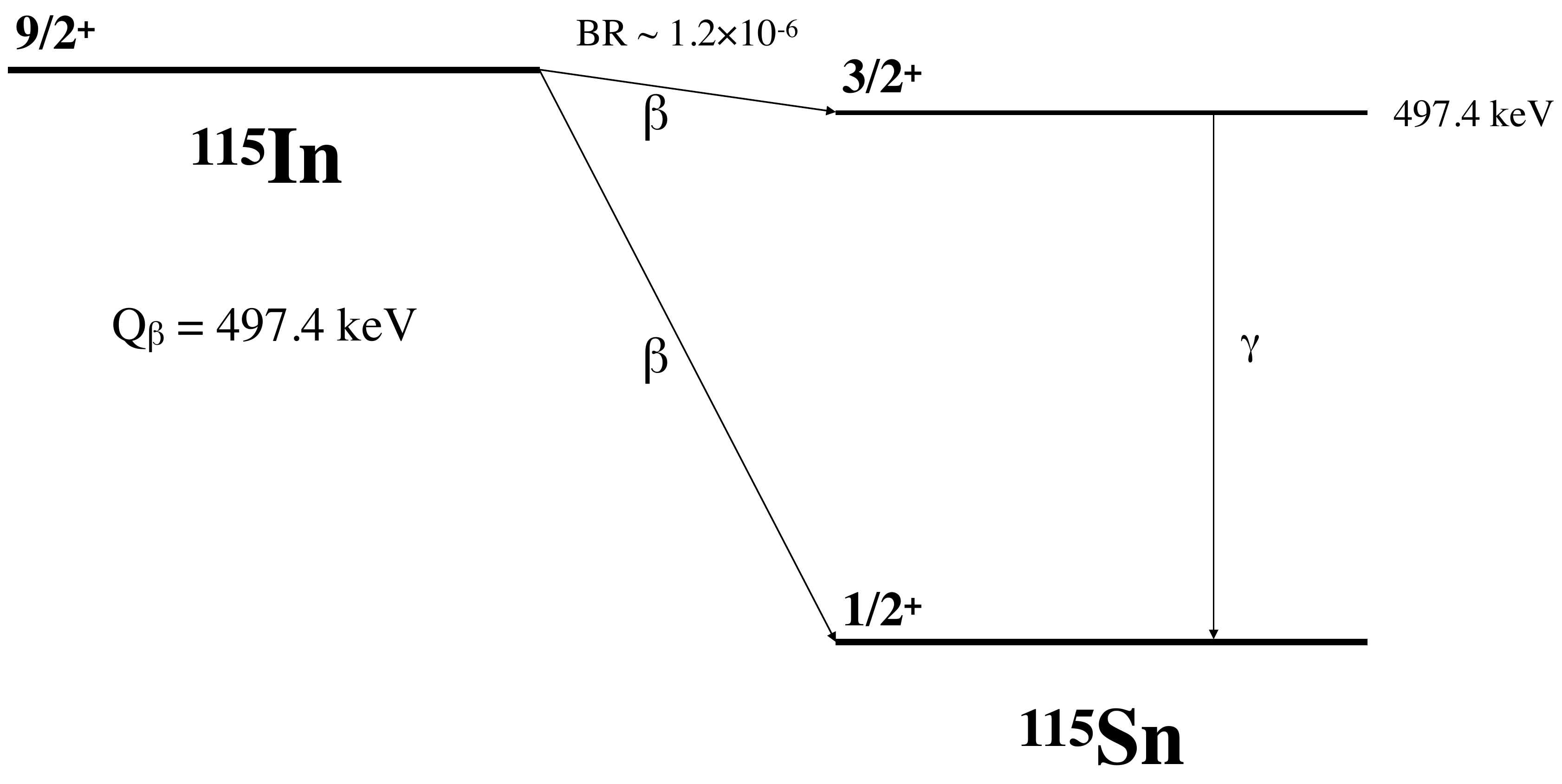}
    \caption{Decay scheme of the $^{115}$In to the ground state (1/2$^{+}$) and excited state (3/2$^{+}$) of $^{115}$Sn. The latter process is candidate to be the $\beta$-decay with the lowest Q-value~\cite{CATTADORI2005333}.}
    \label{fig:DecayScheme}
\end{figure}

$^{115}$In records only two attempts with scintillating liquid loaded by In~\cite{PhysRev.122.1576,PhysRevC.19.1035}, and LiInSe$_{2}$ cryogenic calorimeters~\cite{Tower:2020rlk}. While the first measurements feature a poor energy resolution and energy threshold, the latter reports very promising results in terms of energy resolution (2.5 keV in the range 0 - 500 keV). We recall that $^{115}$In is particularly interesting, because according to its decay scheme, reported in Fig.~\ref{fig:DecayScheme}, it is candidate to feature the $\beta$-decay with the lowest Q-value~\cite{CATTADORI2005333}.

Forbidden decays have always aroused interest, as they represent a benchmark for nuclear models that predict their spectral shape and half-life. Since these decays feature a high transferred momentum, such as {\it neutrinoless double-beta decay}, they represent also an attractive tool to investigate the nuclear aspects of this transition. A detailed review of theoretical and experimental achievements in this field can be found in Ref.~\cite{EJIRI20191}.
Among the most powerful technique for this study, low-temperature detectors based on Metallic Magnetic Calorimeter were demonstrated to be an excellent tool to perform a systematic study of $\beta$-spectrum shapes. They are actually limited by bremsstrahlung within the metal absorber and electron back-scattering, which could introduce artefacts in the spectra, thus preventing a reliable spectral shape evaluation for some specific decays~\cite{LOIDL2019108830}. The same issues affect also the most advanced tool for electron spectroscopy, i.e. Silicon Drift Detector~\cite{Gugiatti:2020wad}, even if recent results demonstrate the possibility to include these effects in the Monte Carlo simulations~\cite{Biassoni:2020oaj}.
On the other hand, cryogenic macro-calorimeters have recently been used for the first time to study the spectral shape of the two-neutrino double-beta decay~\cite{Azzolini:2019yib,Armengaud:2019rll,Adams:2020dyu},demonstrating their potential for this application.
From this awareness comes the idea of dedicating a specific effort to the measurement of the spectral shape of the rare beta decay with cryogenic calorimeters. Nevertheless, the required sensitivity could be achieved only with proper target material selection.
In this work, we propose an innovative method to address $^{115}$In $\beta$-decay spectral-shape studies based on low-temperature macro-calorimeters.
In section \ref{Sec:Xtal} we describe the crystal preparation, in section \ref{Sec:HPGe} we present the results on internal contaminations obtained with $\gamma$-spectroscopy, while in section \ref{Sec:Measurements} we discuss the preliminary measurement of In$_2$O$_3$ as material for a cryogenic calorimeter. Finally, in section \ref{Sec:Conclusions} we describe the next steps to carry out a precise measurement for spectral shape studies.

\section{Crystal Growth and Sample Preparation}\label{Sec:Xtal}
An In$_2$O$_3$ crystal sample for the present study of size $5\times5\times4$~mm$^3$ was prepared from a bulk single crystal grown directly from the melt within a thin iridium crucible by a novel technique ``Levitation-Assisted Self-Seeding Crystal Growth Method'', that was defined and developed by Galazka et al.~\cite{GALAZKA201461,GALAZKA2013349}.
For the growth 4N purity In$_2$O$_3$ powder was used, which was dried at 900~$^o$C in air for 10 h to remove possible moisture and carbonates. 
In this method, the In$_2$O$_3$ starting material is subjected to a controlled decomposition that induces its high electrical conductivity which couples with the electromagnetic field of the induction coil heating the iridium crucible. 
Such coupling generates an extra heat source (in addition to radiation and conduction) that facilitates melting and levitation of a portion of the liquid In$_2$O$_3$. 
By levitating a portion of the melt, a liquid neck between two liquid regions is formed, acting as a seed during solidification. 
Cooling such melt structure down leads to solidification of the melt on both sides of the seed in the form of single crystals of 35 mm in diameter and about 7 mm thick. 
The sample with oriented (111) planes was prepared from the obtained single crystal by cutting and polishing, and by subsequent thermal treatment at 800 C in O$_2$ for 20 h to significantly reduce the density of In nano-particles created by enhanced decomposition during crystal growth at high temperatures $\geq$ 1950~$^o$C and subsequent cooling down~\cite{C2CE26413G,Albrecht2014}. 
It should be noted that at high temperatures In$_2$O$_3$ undergoes extensive thermal decomposition producing several volatile species with O$_2$ and In$_2$O$_3$ having the highest partial pressures~\cite{GALAZKA201461,GALAZKA2013349}. 
As a result, formed oxygen vacancies precipitate in the crystals during cooling down and create adjacent In nano-particles which are visible as inclusions of a typical size 2 – 100 nm and density of 10$^{11}$ – 10$^{15}$ cm$^{-3}$~\cite{Albrecht2014}.
The scattering at such nano-particles makes the as-grown crystals dark (e.g.~brown to black). 
Annealing in the presence of oxygen reduces the density and size of the In nano-particles in the crystals to values of about 10$^{11}$ cm$^{-3}$ and 2 nm, respectively~\cite{Albrecht2014}. 
After annealing, the In$_2$O$_3$ crystals become yellowish and fully transparent in the visible spectrum. 
The In$_2$O$_3$ crystal sample was semiconducting with the free electron concentration and electron mobility of $4\times10^{17}$~cm$^{-3}$ and 160 cm$^2$V$^{-1}$s$^{-1}$, respectively, according to Hall effect measurements.

\section{Spectroscopic measurement with HPGe detector}\label{Sec:HPGe}

We investigate the internal contaminations of In$_2$O$_3$ crystal, performing a $\gamma$-spectroscopy measurement in the STELLA (Sub-Terranean Low-Level Assay) facility at LNGS (Laboratori Nazionali del Gran Sasso) of INFN (Assergi, Italy). A detailed description of the experimental setup can be found in Ref.~\cite{Laubenstein:2017yjj}.
We place the crystal into the well of an ultra low-background high purity germanium (HPGe) well-type detector. This kind of detector is optimised for a very small sample, having a 35.2\% efficiency relative to a 3$\times$3 in NaI(Tl) crystal scintillator and a thin 0.75 mm aluminium window. The detector is surrounded by a composite shield starting on the outside with 10 cm low-activity lead ($<$100 Bq/kg of $^{210}$Pb), followed by another 5 cm of even lower activity lead ($<$6 Bq/kg of $^{210}$Pb) and then 5 cm of oxygen-free high conductivity copper. Finally, the shield and detector are enclosed in an air-tight housing kept at slight over-pressure and continuously flushed with boil-off from liquid nitrogen to prevent and remove radon gas from the setup.

We report in Tab.~\ref{tab:HPGe} the list of internal radioactive nuclides investigated. These values are obtained using the procedure presented in Ref.~\cite{HEISEL2009741}. We found no evidence of any daughter nuclides from the natural decay chains of $^{235}$U, $^{238}$U, and $^{232}$Th, so that we set upper limits ranging from few Bq/kg up to tens mBq/kg. We also report the limits on the activity of other commonly observed nuclides, in particular for $^{40}$K, from natural radioactivity, and artificial $^{137}$Cs.

\begin{table}[!ht]
    \centering
    \caption{Activity of internal radioactive contaminations in the In$_2$O$_3$ crystal measured with a HPGe detector. Values are present in units of Bq/kg, and limits are at 95\% C.L. We evaluate the limit on $^{210}$Pb using the bolometric measurement presented in this paper, assuming it produces all the events above 600 keV via $^{210}$Bi decay.}
    \begin{tabular}{lrr}
         Chain	& Nuclide & Activity\\
                &         & [Bq/kg]\\
                \hline
$^{232}$Th	&$^{228}$Ra	&$<0.18$ \\
            &$^{228}$Th	&$<0.12$ \\
$^{238}$U	&$^{234}$Th	&$<1.7$ \\
            &$^{234m}$Pa&$<7.4$ \\
            &$^{226}$Ra	&$<0.077$ \\
            &$^{210}$Pb	&$<3.5$ \\
$^{235}$U	&$^{235}$U	&$<0.072$ \\
            &           &       \\
            &$^{40}$K	&$<2.0$ \\
            &$^{137}$Cs	&$<0.042$ \\
    \end{tabular}
    \label{tab:HPGe}
\end{table}

\section{Experimental Setup and Measurement}\label{Sec:Measurements}
\begin{figure}
    \centering
    \includegraphics[width=0.45\textwidth]{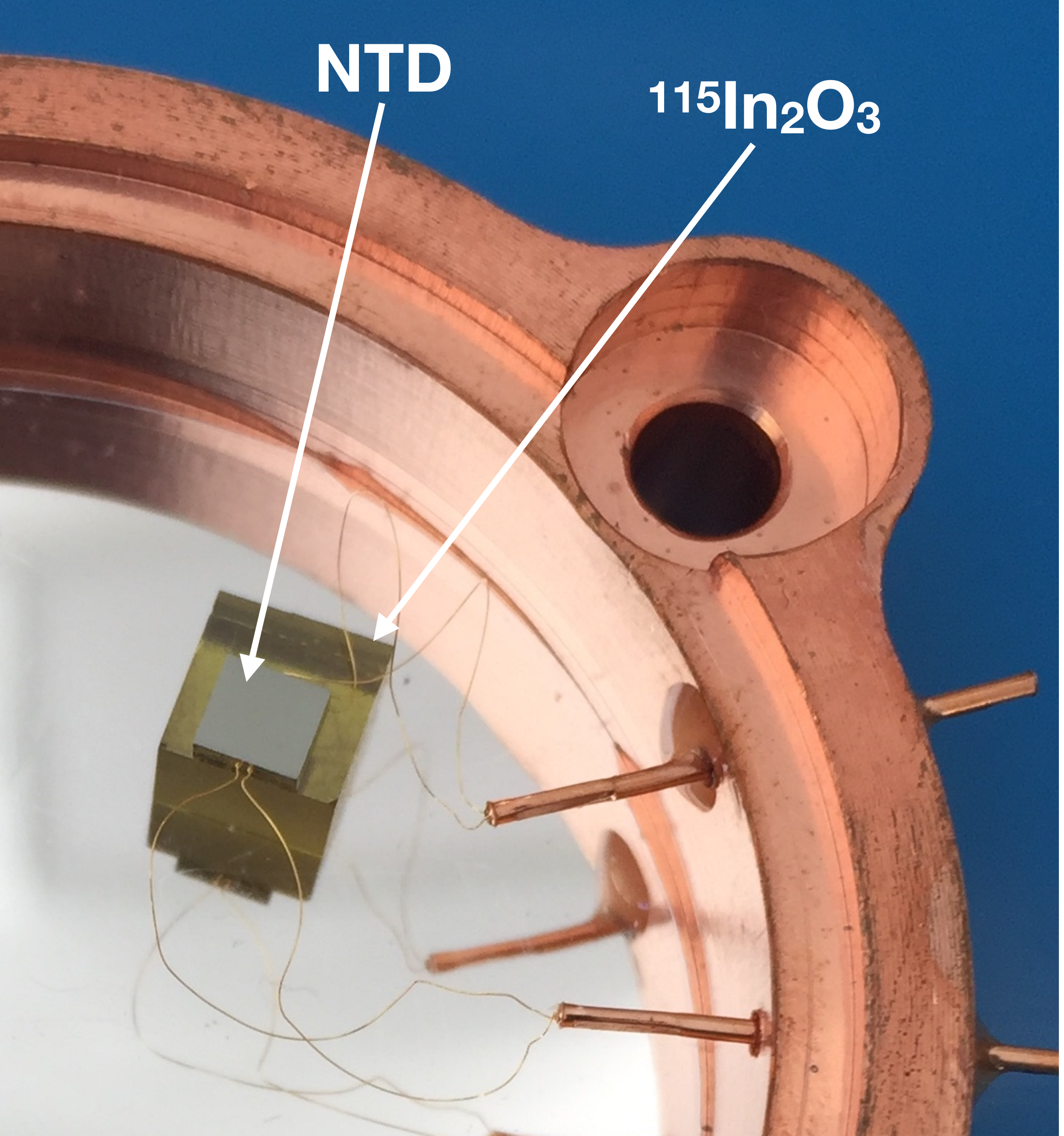}
    \caption{Detector setup used to measure the In$_2$O$_3$ crystal as cryogenic calorimeter. We glued on the crystal a germanium NTD sensor, whose signal is readout using 50 $\mu$m gold wires.}
    \label{fig:1}
\end{figure}

\begin{figure}
    \centering
    \includegraphics[width=0.5\textwidth]{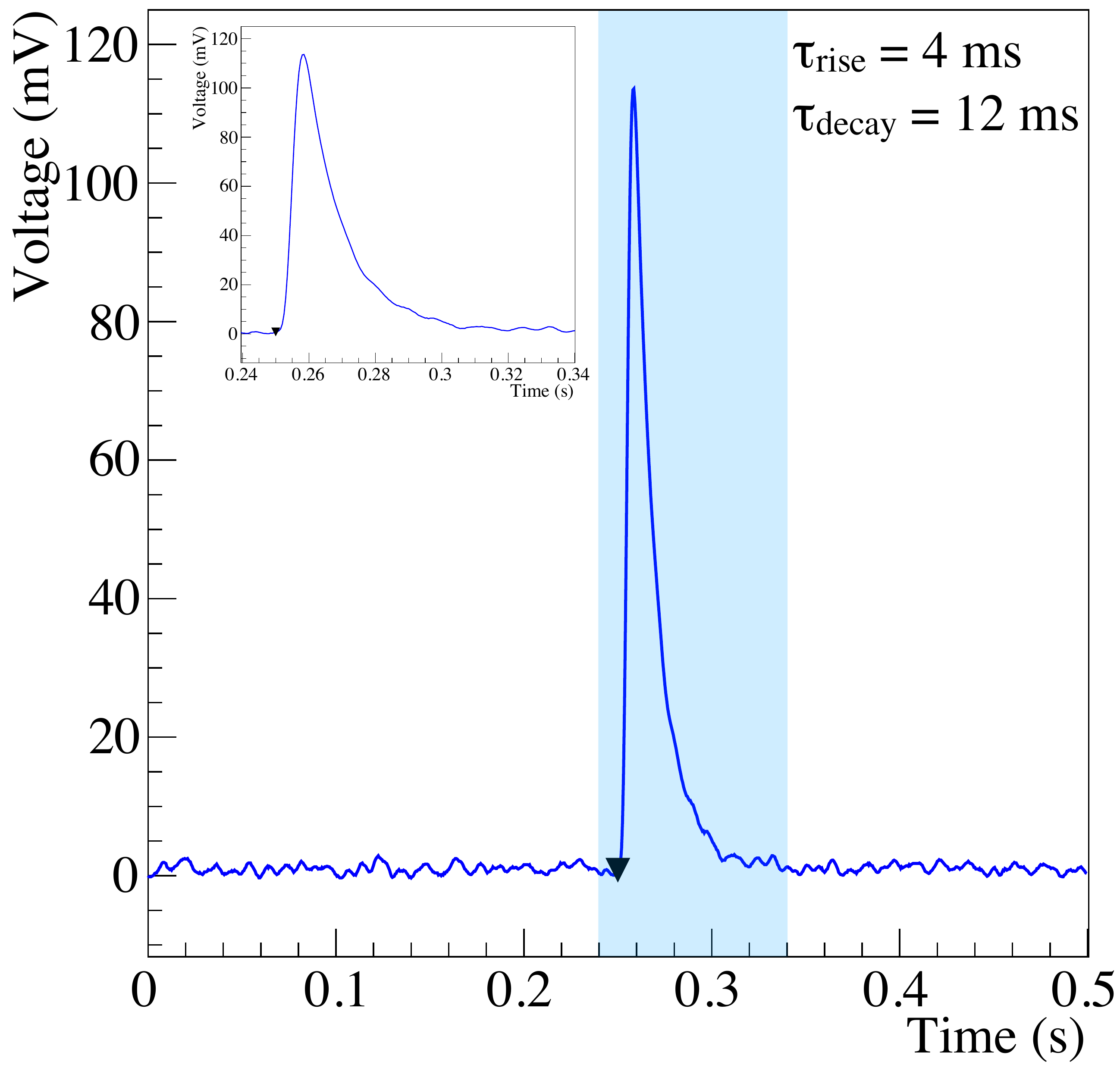}
    \caption{Typical signal of In$_2$O$_3$ detector, featuring a rise time of 4 ms and a decay time of 12 ms. The whole signal is fully contained in 100 ms, as shown in the inset.}
    \label{fig:InoPulse}
\end{figure}

We equip the In$_2$O$_3$ crystal with a Neutron Transmutation Doped (NTD) germanium thermistor~\cite{Haller}, acting as temperature-voltage transducer.
We mount the detector in a copper holder, where the crystal stands on a plastic reflective foil as show in Fig.~\ref{fig:1}.
The entire setup was enclosed in a Cu box and thermally coupled to the mixing chamber of the CUPID R$\&$D cryostat, a $^3$He/$^4$He dilution refrigerator installed deep underground in Hall C of the Laboratori Nazionali del Gran Sasso, Italy. 
To avoid vibrations reaching the detectors, the box is mechanically decoupled from the cryostat exploiting a two-stage pendulum system~\cite{Pirro:2006mu}.
The thermistor is biased with a quasi-constant current produced by applying a fixed voltage through large (27+27 or 2+2 G$\Omega$) load resistors~\cite{Arnaboldi:2017aek}, operating the sensor at a base resistance R$_b = 7.2$ M$\Omega$, and a working resistance of  R$_w = 3.4$ M$\Omega$.
When a particle releases energy in the crystal, a thermal pulse is produced which is subsequently transferred to the NTD sensor, changing the resistance of the thermistor. 
This, in turn, creates a voltage change across the current-biased NTD which is amplified using front end electronics located just outside the cryostat~\cite{Arnaboldi:2017aek}. The signals are then filtered by an anti-aliasing 6-pole Bessel filter (with a cutoff frequency of 100 Hz) and finally fed into a NI PXI-6284 18-bit ADC.
The sampling rate of the ADC was 2 kHz. A detailed description of the DAQ system can be found in Ref.\cite{DiDomizio:2018ldc}. 
The amplitude and the shape of the voltage pulses are then determined via off-line analysis, following the data processing technique detailed in Ref.~\cite{Azzolini:2018yye}. In particular, the pulse amplitude of the thermal signals is estimated by the Optimum Filtering technique~\cite{Gatti:1986cw}, maximising the signal-to-noise ratio in a way that improves the energy resolution and lowers the threshold of the detector. We report in Fig.~\ref{fig:InoPulse} the typical signal observed with In$_2$O$_3$ detector. Being the dimensions of the In$_2$O$_3$ crystal very small (0.1~cm$^3$) and the time measurement very short (34 hours), the only visible $\gamma$-line in the spectrum is the $e^+e^-$ annihilation peak at 511~keV, on which we perform both the Thermal Gain Stabilization (TGS) and energy calibration. 

\begin{figure}[t]
    \centering
    \includegraphics[width=0.48\textwidth]{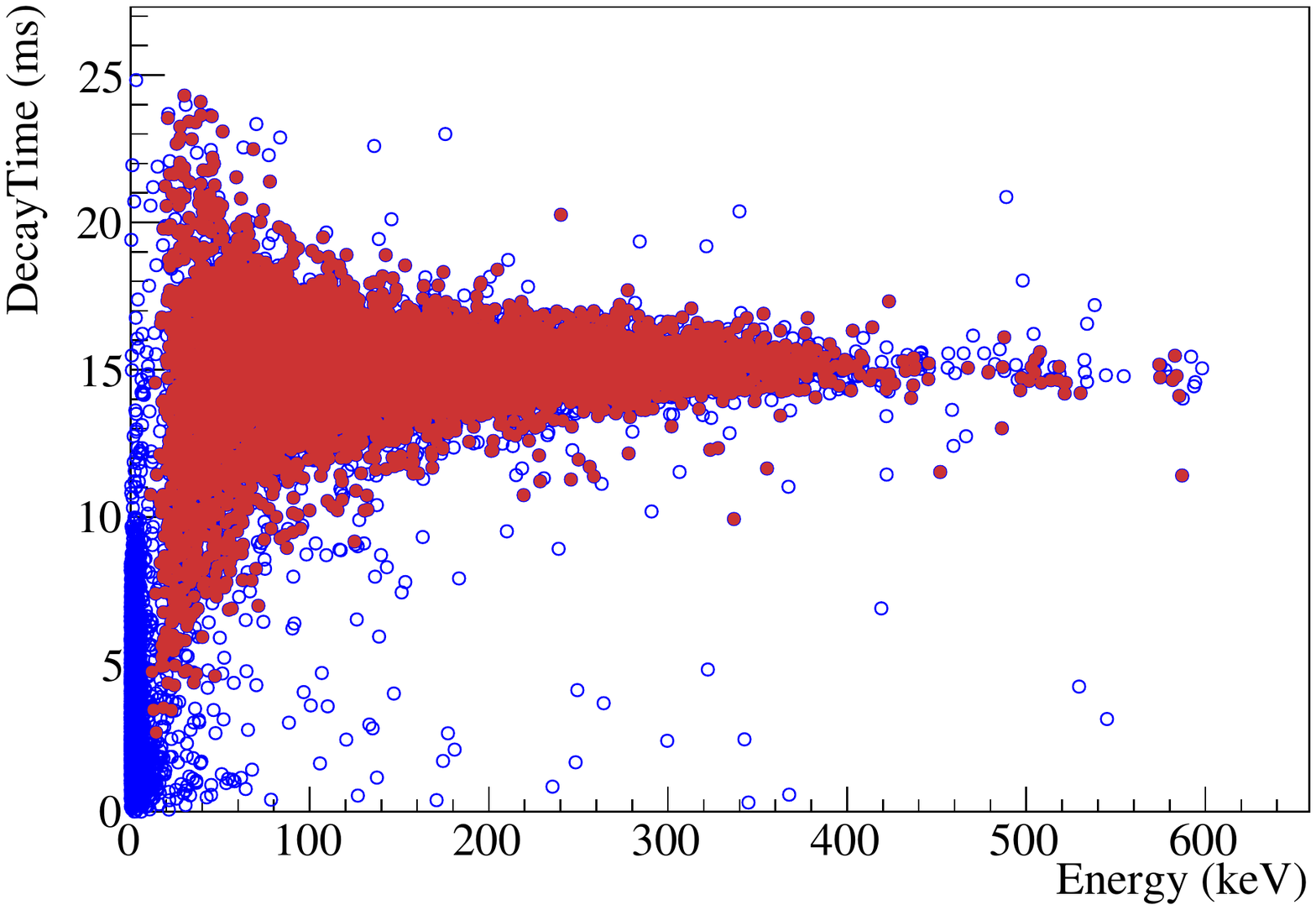}
    \caption{Decay time of the signal pulse as a function of the energy released in the crystal, without (blue) and with (red) quality selection cut. At low-energy, we can clearly identify two populations (i.e. signal vs. noise), whose separation threshold is $\sim20$ keV.}
    \label{fig:InoScatterPlot}
\end{figure}

For each triggered pulse, we define also the following variables: 
\begin{itemize}
    \item number of triggered signal within the acquisition window;
    \item rise-time as the time difference between the 90\% and the 10\% of the leading edge;
    \item decay-time as the time difference between the 30\% and 90\% of the trailing edge.
\end{itemize}
In order to remove spurious events, we select single-triggered waveforms, with a rise-time and a decay-time within a $\pm 1.5 \sigma$ interval centred in their average value. The result of the selection is depicted in Figs.~\ref{fig:InoScatterPlot} and ~\ref{fig:InoSpectrum}.

\section{Results and Perspectives}
\begin{figure}
    \centering
    \includegraphics[width=0.48\textwidth]{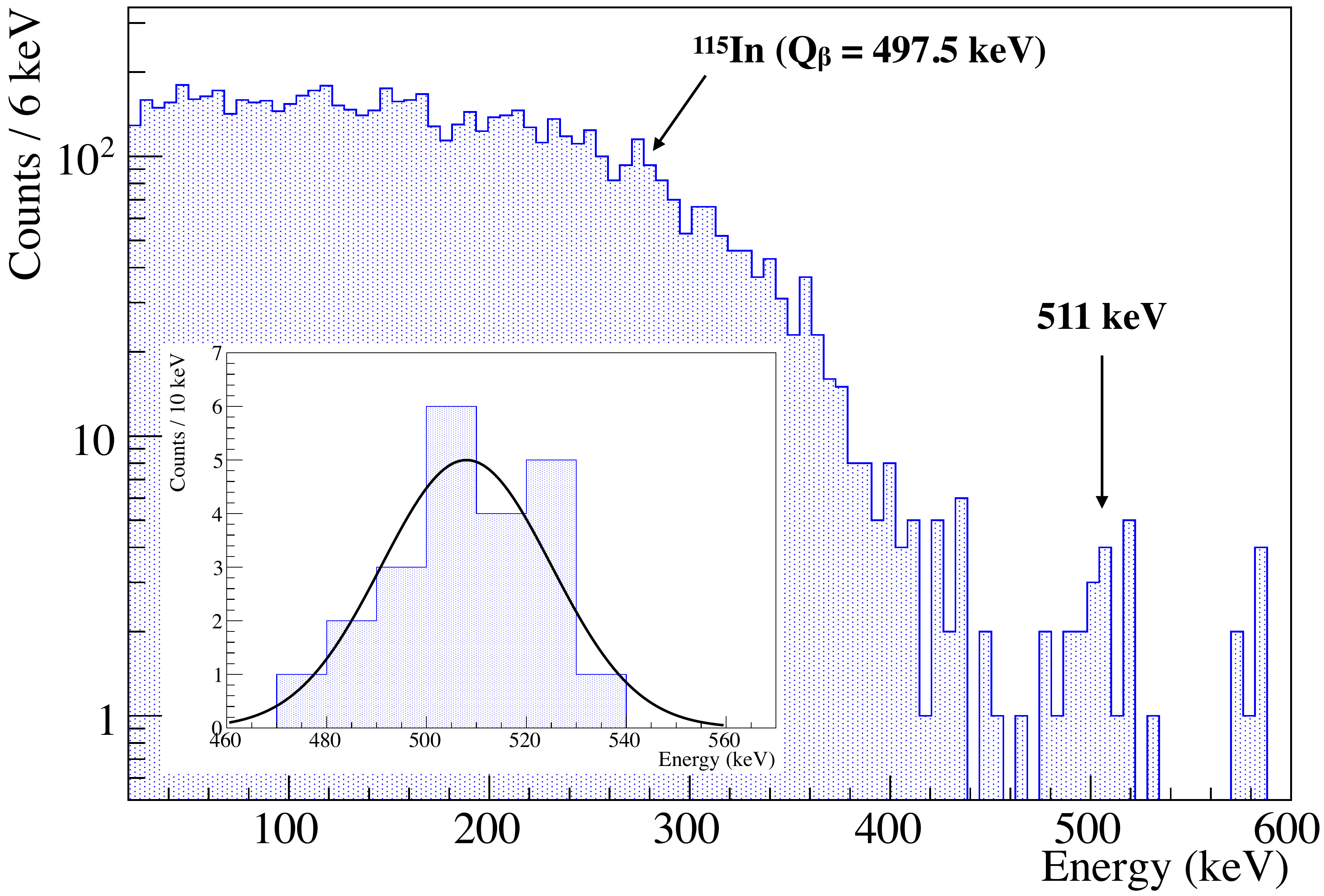}
    \caption{Energy spectrum obtained from the preliminary measurement of the In$_2$O$_3$ crystal (34 hours). Even if data quality is not satisfactory for spectral shape studies, the $^{115}$In $\beta$-decay clearly dominates the collected data, reflecting the very-low background achievable with the proposed technique. We report in the inset the Gaussian fit on the 511~keV calibration peak, whose energy resolution is $(17 \pm 4)$ keV.}
    \label{fig:InoSpectrum}
\end{figure}

Even if the measurement conditions are not optimal, we note that the $^{115}$In $\beta$-decay dominates the spectrum in Fig.~\ref{fig:InoSpectrum}, thus the signal-to-background ratio is very promising. In particular, we can observe that above 50 keV the good signals (red dots in Fig.~\ref{fig:InoScatterPlot}) are a factor 60 higher than the spurious events removed by the quality cuts. The energy resolution at 511 keV is $\sigma = (17 \pm 4)$ keV, which is sub-optimal due to a not-effective TGS performed on a low-statistics peak.
From the scatter plot in Fig.~\ref{fig:InoScatterPlot}, we can get a rough estimate of the signal threshold of around 20 keV, while for such a small crystal a threshold lower than 1 keV is expected. Also this effect can be traced back to the non-optimal measurement conditions, and to an intrinsic limit due to the large thermal capacity of the sensor with respect to the crystal.
Given the features of the crystals and a detection efficiency of 76.1\% for the $^{115}$In $\beta$-decay, the expected signal rate is 85~mHz. This value is compatible with the actual observed rate of 105~mHz.
Being the signal fully contented in 100~ms (Fig.~\ref{fig:InoPulse}), we can shrink the acquisition window from 500~ms at this level, thus reducing the pile-up probability. This allows us to choose a bigger crystal (e.g.~1~cm$^{3}$), increasing both the source activity (0.95~Hz) and the containment efficiency (87.3~\%), while keeping the pile-up probability below the 10~\%.
Given the promising results obtained in this preliminary test, we are planning a second measurement with an optimized experimental setup. We will equip the detector with an silicon heater, in order to exploit the standard TGS technique and perform a pulse scan to determine the low energy threshold very precisely.
Also the calibration can be remarkably improved, using a removable $\gamma$-source made by a thoriated tungsten wire, which provides X-rays and $\gamma$-peaks ranging from few keV up to 2.6 MeV~\cite{Nagorny:2019syb}.
In this condition, we expect to reach the low energy threshold at the level of 1 keV along with an improved energy resolution at order of 5 keV (FWHM) on the 511 keV peak. To improve the detector performance, we plan in the future to use, once prepared, an In$_2$O$_3$ single crystal of a larger volume and reduced density of In nano-particles.

\section{Conclusions}\label{Sec:Conclusions}
In this work, we propose a novel approach to measure the spectral shape of $^{115}$In $\beta$-decay with a cryogenic calorimeter based on a In$_2$O$_3$ crystal. Despite the sub-optimal working conditions, this preliminary test demonstrates the effectiveness of this approach in terms of detection efficiency, signal-to-noise ratio and low-background. In order to improve the detector performance, we will equip the crystal with a smaller size NTD, and an heater providing reference pulses for thermal gain stabilisation. Moreover, we will exploit a removable thorium/tungsten calibration $\gamma$-source close to the crystal for the energy calibration. 
Since In$_2$O$_3$ crystal is a semiconductor with a high concentration of electrons and high mobility, we consider depositing aluminium contacts and utilise Neganov-Luke-Trofimov amplification of the phonon signal. It will lead to reducing energy threshold and significantly improving energy resolution. The new experimental setup will allow us to carry out a longer and more stable measurement in order to study the spectral shape of the $^{115}$In $\beta$-decay~\cite{Kostensalo:2017jgw}. Studies of this type are encouraged to clarify some crucial aspects of nuclear physics, such as the quenching of axial coupling constant ($g_A$) in the nuclear medium, which have important implications also in astroparticle physics~\cite{EJIRI20191}.

\section{Acknowledgements}
Crystal growth of In$_2$O$_3$ was partly performed in the framework of GraFOx, a Leibniz-Science Campus partially funded by the Leibniz Association, Germany. The authors express their gratitude to Dr. Thomas Teubner from the Leibniz-Institut für Kristallzüchtung for critical reading of the manuscript.

\bibliography{main}
\bibliographystyle{spphys} 

\end{document}